# NEW APPROACHS TO QUANTUM COMPUTER SIMULATON IN A CLASSICAL SUPERCOMPUTER

J. Robert Burger
California State University Northridge

**ABSTRACT**
Classical simulation is important because it sets a benchmark for quantum computer performance. Classical simulation is currently the only way to exercise larger numbers of qubits. To achieve larger simulations, sparse matrix processing is emphasized below while trading memory for processing. It performed well within NCSA supercomputers, giving a state vector in convenient continuous portions ready for post processing.

## 1. BACKGROUND

Simulation is critical to confidence and to the yet-to-be-designed quantum computer, as it was for nuclear energy. Certainly simulation, scaled as it may be, will be less costly than experimental testing of a quantum computer. Without extensive simulation, a working model invariably will have unacceptable limitations, such as low probability for a correct answer. Today, better simulators, to the limits of classical computational power, are urgently needed to incorporate recent results in reversible logic, and to study performance in space and time [1,2,3].

There have been serious efforts at simulation in the private sector [4]. A beautiful simulator is Universal Quantum Computation Simulator, listed on the web in 1999, with a free version available to the public [5]. This version accepts only a limited number of qubits however.

Another interesting package entitled LGP2 and QC Simulator was listed on the web in 1997. This simulator can employ 'genetic' programming system to discover new quantum algorithms [4]. But the overhead involving random variables is likely to subtract from the power available for simulation in a conventional computer.

Attempts at simulation have emerged in studies of quantum neural networks [6]. Here the state of a quantum system is computed in the time domain, in contrast to the work below, which considers only the steady state.

A general comprehensive reference that defines terms and ideas in quantum computation and information is readily available [7]. To begin, a quantum computer simulator, like the physical implementation itself, must be initialized [8,9,10,11]. Below we allow arbitrary magnitude and phase in a state vector to begin.

In a basic model with M qubits, M bits can be used to address a given entry in the state vector. Considering projected numerical requirements, and considering redundancy for error correction, the ideal number of address bits should be perhaps a few hundred [1]. Jobs of this size run classically only for special cases, that is, with limited quantum parallelism [12]. Assuming full parallelism and M = 64, the word size in a modern workstation, it is impossible today to store completely even one state vector ($2^M \approx 1.8 \times 10^{19}$) in RAM. The first question is how <u>large</u> a state vector can we run? The second question is how <u>long</u> will it take?

To partially answer such questions in today's technology, the author implemented his sparse matrix algorithm using modi4 at NCSA (National Center for Supercomputer Applications). This system offers 64 processors, 16 Gb, and 123 Gb of scratch disk.

## 2. SPARSE MATRIX APPROACH

### (a) Overview

In this presentation there is a unitary matrix that transforms the initial state $\mathbf{k_1}$ into a new state $\mathbf{k_2}$. Any matrix that satisfies $\mathbf{U^+U = I}$ is unitary, for example,

$$\begin{bmatrix} & & & 1 \\ 1 & & & \\ & & 1 & \\ & -1 & & \end{bmatrix}$$

To simplify, we assume a class of unitary matrices implied by a wiring diagram. In contrast to the above matrix, any one step in a wiring diagram gives a matrix that is symmetrical about the principle diagonal as shown below. Assuming that phase gates are not used, each matrix element produced by a wiring diagram will be non-negative. Sparse matrices from wiring diagrams have only one non-zero entry per row and it is unity. Below in parts (b) through (g) it is shown how to calculate a unitary matrix from a wiring diagram.

### (b) Approach When a(1) is the Target Bit

In this paper the identifier, or address bits a(M)… a(2)a(1) where a(1) is the LSB in Figure 1, point to an element in the state vector. In Step 1, a(1) is the target bit and a(2) is the controlling bit. The steps in the wiring diagrams, of course, scramble addresses in a certain useful way. Assuming eight addresses, each is modified, for example, 011 to 010.

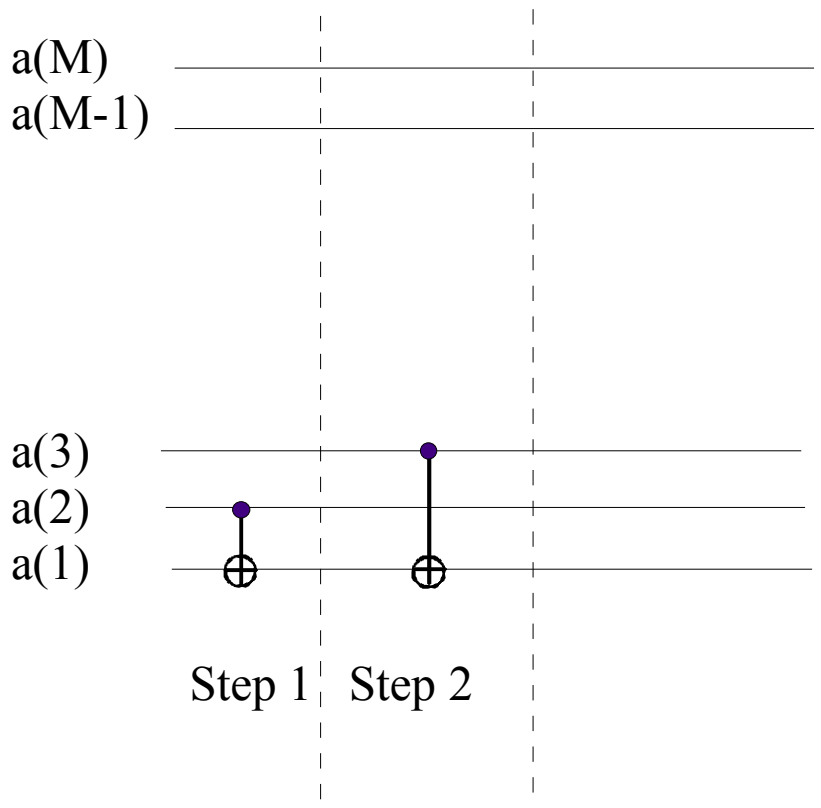

**Figure 1. Example Wiring Diagram**.

Looking ahead a little, the associated unitary matrix can be constructed beginning with the identity matrix **I**. At the end of L Steps, $\mathbf{k_2} = \mathbf{U_L}\, \mathbf{k_1}$, where $\mathbf{U_L}$ may be thought of as the product of the individual unitary matrix for each step. For future reference, the elements of $\mathbf{k_1}$ are addressed from top to bottom in binary from 0 to $2^{M-1}$. The elements of $\mathbf{k_2}$ are also addressed from top to bottom in binary from 0 to $2^{M-1}$.

The first step in the wiring diagram produces $\mathbf{U_1}$ determined as follows: Imagine a binary count from 0 to $2^{M-1}$ beginning at the top of a resulting state vector. In this count, when a(2) reaches its first pairs of 1s, a corresponding 2x2 sub matrix is affected. The matrix toggles from an identity matrix to a 'reversing' unitary matrix as shown:

$$\begin{bmatrix} 1 & & & & & & & \\ & 1 & & & & & & \\ & & 0 & 1 & & & & \\ & & 1 & 0 & & & & \\ & & & & 1 & & & \\ & & & & & 1 & & \\ & & & & & & 0 & 1 \\ & & & & & & 1 & 0 \end{bmatrix}$$

Assume that the input is: $\mathbf{k_1} = |S|\,(1\ {-1}\ 1\ {-1}\ 1\ {-1}\ 1\ {-1})'$, that is, 1s and –1s alternating at a certain basic frequency along the vector (prime denotes transpose). |S| is a scalar $|S| = 1/\sqrt{(2^M)}$, a standard normalization that is assumed to be fixed to simplify the presentation without significant loss of generality. The resulting state vector after Step 1 becomes: $\mathbf{k_2} = |S|\,(1\ {-1}\ {-1}\ 1\ 1\ {-1}\ {-1}\ 1)'$. Clearly the frequency of the alternations has changed, which is very important information, although interpretation of this data is for post processing.

To define a cyclic notation, let $\mathbf{k_{1address}}$ hold the integer addresses of $\mathbf{k_1}$; if $\mathbf{U_1}\, \mathbf{k_{1\,address}}$ differs from $\mathbf{k_{1address}}$ an important interchange event has taken place. Interchanges are pairs ($\mathbf{k_{1address}}$, $\mathbf{U_1}\, \mathbf{k_{1address}}$), for example (2,3), (6,7). The use of cyclic notation hides magnitude and phase information, but it succinctly expresses interchanges within the state vector [3].

Returning to sparse matrix processing, a row array x(i) may be defined that contains the locations of the ones. What x(i) holds is the present column number for the entry in row i. The matrix $\mathbf{U_1}$ is represented by a temporary array y(i) for $1 \le i \le 2^M$. It will hold the column number for the entry in row i. It is convenient to declare a variable 'from' represents the controlling bit and a variable 'goto' represents the target bit.

*initialize*
*for all 'goto' and for all i*
*y(i) = i*
*x(i) = i*
*t(goto, i) = 1*

Also initialized above is an optional temporary toggle array t(goto, i) to keep track of whether or not a sub matrix is toggled, where $1 \le i \le 2^M$. What t(goto, i) holds is -1 if the sub-matrix is toggled, and +1 if the sub-matrix is not toggled, to be used later. Toggling refers to the switch between the 2x2 identity and the reversing matrix above. In some applications toggling is useful.

Using replacement statements, consider Step 1 in Figure 1, which shows 'from' = 2 and 'goto' = 1. Consider a binary count of the addresses a(M-1)… a(2)a(1). Pairs of ones in a(2) occurs for i = 3, 4, 7, 8 and so on. For each pair the following is implemented:

*y(i) = y(i)+t(goto, i)*
*y(i+1) = y(i+1)- t(goto, i)*
.
.
.

When 'from'=3 and 'goto'=1, ones in a(3) occurs in quadruples, that is, as two pairs:

*y(i) = y(i) + t(goto, i)*
*y(i+1) = y(i+1) - t(goto, i)*
*y(i+2) = y(i+2) + t(i+2)*
*y(i+3) = y(i+3) - t(i+2)*
.
.
.

Generally there will be a block of ones of size $2^{from-1}$ in which y(i) is toggled as above for each pair of ones in the block.

The structure of a binary count is predictable, so it is possible to perform the toggling efficiently and compactly (without conditional statements). The author calls this process 'intelligent looping' in which only those sub-matrices that are to be toggled are addressed. Without looking at zeros, the program looks only at blocks of ones to perform the toggles according to the wiring diagram.

After a Step, the arrays are initialized: *for all i: x(i) = y(i), and y(i) = 1.* The state vector, if wanted, is available:

*for all i*
*$k_2(i) = k_1(x(i))$*

**(c) a(2) is the Target Bit controlled by a(from), assuming 'from' > 2**
The case 'goto' = 2, or a(2) the target (not shown in the figure) involves a toggle within a unitary 4x4 sub matrix composed of 2x2 sub sub-matrices. For example, consider 'from' = 3 and 'goto' = 2. When the address has a(3) = 1 the following toggle occurs:

$$\begin{bmatrix} 1 & & & & & & & \\ & 1 & & & & & & \\ & & 1 & & & & & \\ & & & 1 & & & & \\ & & & & & & 1 & \\ & & & & & & & 1 \\ & & & & 1 & & & \\ & & & & & 1 & & \end{bmatrix}$$

For each group of ones in a(3), there occurs a quadruple operation:

*y(i) = y(i) + 2 t(goto, i)*
*y(i+1) = y(i+1) + 2 t(goto, i)*
*y(i+2) = y(i+2) - 2 t(goto, i)*
*y(i+3) = y(i+3) - 2 t(goto, i)*

The calculation can be generalized to any group involving multiples of four ones in 'from' > 2. The case 'from' < 2 means that the controlling bit is below the target wire, and is considered as a separate case.

**(d) General Case a(goto)**
Following the above methods in which 'goto' = 1 with sub matrices of 2x2, and 'goto' = 2 with sub matrices of 4x4, the general case of 'goto' = d affects sub matrices of size $2^d$. Within this sub matrix are sub matrices of size $2^{d-1}$ that can be interchanged. Toggling occurs when blocks of ones of size $2^d$ are encountered in the binary counts.

**(e) Unconditional NOT Operation**
In this case each and every appropriate sub matrix is toggled unconditionally. If 'goto' = 1 is the target, for example, then all 2x2 matrices are toggled.

**(f) 'from' < 'goto'**
When the controlling bit is below the target wire, there is little change in the above formulas, except for the locations where ones occur. For example, if wire 1 controls wire 2, the matrix toggles as follows:

$$\begin{bmatrix} 1 & & & & & & & \\ & & & 1 & & & & \\ & & 1 & & & & & \\ & 1 & & & & & & \\ & & & & 1 & & & \\ & & & & & & & 1 \\ & & & & & & 1 & \\ & & & & & 1 & & \end{bmatrix}$$

Every place where the ones occur in the state count, row position is either increased by 2 or decreased by 2 as shown.

**(g) Double Controlled NOT Gates**
A double controlled gate requires processing when the address bits for two controls are true. This is determined by analysis of a binary count and intelligent looping. The extension to multiple controlled NOT gates is possible but beyond the scope of this paper.

**(h) The Second and Subsequent Steps in the Wiring Diagram**
The equivalent of sparse matrix multiplication could be accomplished by calculating the state vector after each step (as above for $\mathbf{k_2}$). Standard sparse matrix multiplication has the advantage that it does not require the state vector [13]. Because the unitary matrix has only one entry per row, an entry of unity, multiplication is efficient.

There are several ways to accomplish sparse matrix multiplication. One of the simplest is to assume the column entries in matrix 1 are organized by row, x(i). Assume also that the column entries in matrix 2 are organized by row, y(i). Then the product matrix (organized by row) is y(x(i)). To see an example of how this works, consider Figure 2.

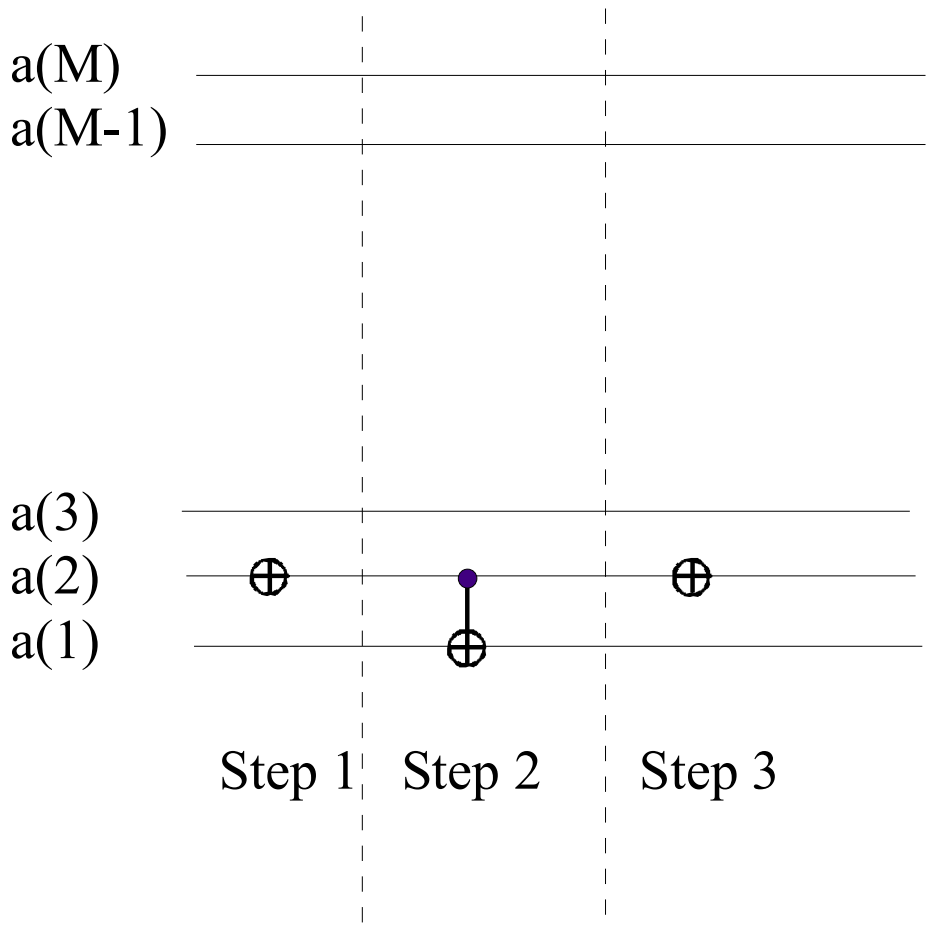

**Figure 2. Example For Multiplication**

Initially x(i) = y(i) = i for all i, where x(i) represents the identity matrix. In Step 1, the matrix $\mathbf{H_1}$ is represented by y(i):
*y(i) = y(i) + 2 t(goto, i) =3*
*y(i+1) = y(i+1) + 2 t(goto, i) = 4*
*y(i+2) = y(i+2) - 2 t(goto, i) = 1*
*y(i+3) = y(i+3) - 2 t(goto, i) = 2*
.
.
.
Let:
*x(1) = y(1) = 3*
*x(2) = y(2) = 4*
*x(3) = y(3) = 1*
*x(4) = y(4) = 2*
.

.
.
Initialize: $y(i) = i$. The $\mathbf{H_2}$ matrix is represented by:
y(1) = y(1) = 1;
*y(2) = y(2) = 2;*
*y(3) = y(3)+t(goto, i) = 4*
*y(4) = y(4)- t(goto, i) = 3*
.
Multiplying:
*x(1) = y(x(1)) = 4*
*x(2) = y(x(2)) = 3*
*x(3) = y(x(3)) = 1*
*x(4) = y(x(4)) = 2*
Next, initialize: y(i) = i. In Step 3 the unitary $\mathbf{H_3}$ is represented (as it was in Step 1) by:
*y(1) = y(1) + 2 t(goto, i) = 3*
*y(2) = y(2) - 2 t(goto, i) = 4*
*y(3) = y(3) + 2 t(goto, i) = 1*
*y(4) = y(4) - 2 t(goto, i) = 2*
.
.
.
The output matrix is:
*x(1) = y(x(1)) = 2*
*x(2) = y(x(2)) = 1*
*x(3) = y(x(3)) = 3*
*x(4) = y(x(4)) = 4*
This is the correct description of the sparse matrix after three steps, and will result in the cycling (0,1), (4,5) etc. in the state vector:

$$\begin{bmatrix} & 1 & & & & & & \\ 1 & & & & & & & \\ & & 1 & & & & & \\ & & & 1 & & & & \\ & & & & & 1 & & \\ & & & & 1 & & & \\ & & & & & & 1 & \\ & & & & & & & 1 \end{bmatrix}$$

**(i) Parallelism**
In a given step there is a given target with a given control. The *intelligent loop* for that control may be broken into parts that run simultaneously. Obviously too many parts, that is, an approach to quantum parallelism requires an excessive number of processors.

An interesting special case occurs in which sequential steps in a wiring diagram can execute in parallel in a simulation. Assume a target wire a(i) identified by the target index i in the above wiring diagram system. A sequence of steps in the above style of wiring diagram for which the next targets satisfy $j \geq i$ is especially easy to process. In this case the associated matrices can be multiplied by toggling sub-matrices. For example, this situation applies when a sequence of steps have the same target. This processing shortcut does not work the other way ($j > i$) as may be seen by analysis of Figure 2. Multiplication must be done the usual way, because the first toggle places entries in the wrong quadrant for shortcut multiplication.

When toggling, be it for intelligent loops, or sequential steps, it is obvious that subprograms can in parallel. For each subprogram the appropriate toggles are retained in an array T(from, goto, i) which is 1 for no toggle, and –1 for toggle. The 'i' indicates the location of the sub-matrix. The toggle status after parallel processing is determined by post processing:
*for all 'from'*

$$t(goto, i) = \prod T(from, goto, i)$$

What t(goto, i) holds is 1 for NO toggle, and –1 for TOGGLE. Array t(goto, i) indicates the toggle status of a sub-matrix of size $2^{goto}$ containing sub sub-matrices of size $2^{goto-1}$. This method takes advantage of the fact that two toggles (-1)(-1) equal no toggle. As a final step, the row array is updated using appropriately applied operations of the form:
$y(i) = y(i) \pm 2^{goto-1}\, t(goto, i)$

Toward the end of a simulation the elements in $\mathbf{k_2}$ need to be calculated. At this point, random elements of $\mathbf{k_1}$ are needed. They flow from a calculation involving elements of a Kronecker (direct matrix) product (Appendix 1).

**(i) Summary of Method**
The author's *sparse matrix algorithm* is summarized below:

- Construct an identity matrix **I** of dimension $2^M$, where M is number of qubits. Represent **I** by x(i) = i and initialize y(i) = i for all i.
- In the first step, a target wire labeled $a_d$, $1 \leq d \leq M$ results in toggling in sub-matrices of dimension $2^d$. Toggling is represented by changes in column positions held by y(i).
- The elements that change in y(i) can be pinpointed by intelligent looping, that is, only 1s in the addresses result in toggles in sub-matrices of dimension $2^d$.
- Let x(i) = y(x(i)) or toggle as appropriate to implement sparse matrix multiplication; then y(i) = i is initialized ready for the next step in the wiring diagram.
- After processing all steps, the output state vector is computed as: $k_2(i) = k_1(x(i))$

## 4. COMPUTATIONAL EXAMPLE

The following special case was implemented on a NCSA system named modi4. The test program in Figure 3 is going to compute a particular quantum parity function of qubits 2 through M. The result is targeted to the bottom qubit. The input state with $2^M$ entries is assumed to be: $k_1 = (1\ {-1}\ 1\ {-1}\ \ldots 1\ {-1})/\sqrt{2^M}$ corresponding to the states |00…01> for the M qubits, with the |1> on the bottom. The program used full arrays for t, $k_1$, and $k_2$. However, address space was observed to be inadequate for M > 31 in the author's setup.

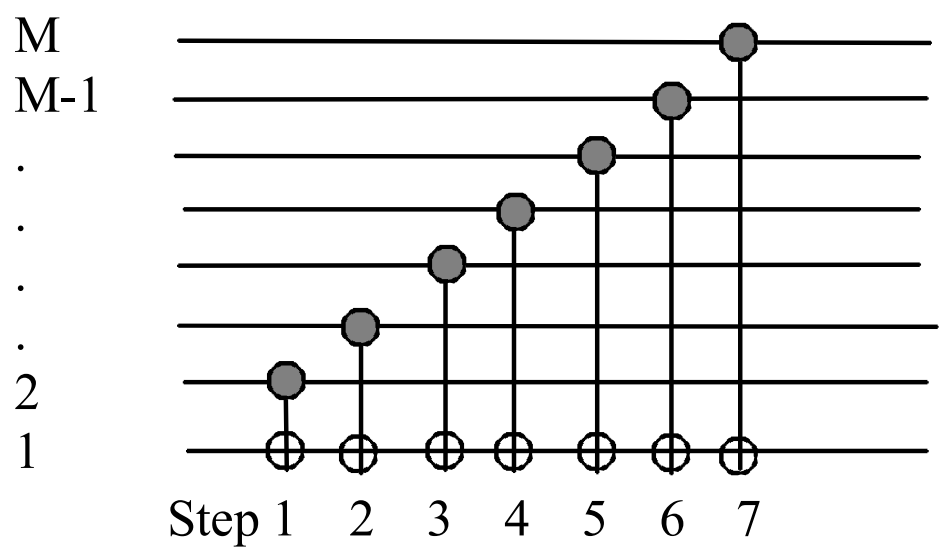

**Figure 3. Test Program.**

Using the automatic parallelizing option (apo) with up to 64 processors Figure 4 resulted. Plotted are cpu times divided by 64, that is, approximate run times. Fewer than 28 qubits does not require all 64 processors, which is why the curve deviates from a straight line on the left. Note the apo could not run nested loops in parallel, although it did mark the inner loop as parallelized. The extrapolated results are sensitive to the slope of the data (roughly x3.4per bit shown going from M = 30 to 31). This slope depends on computer efficiency with large arrays, so extrapolation to large M is not recommended.

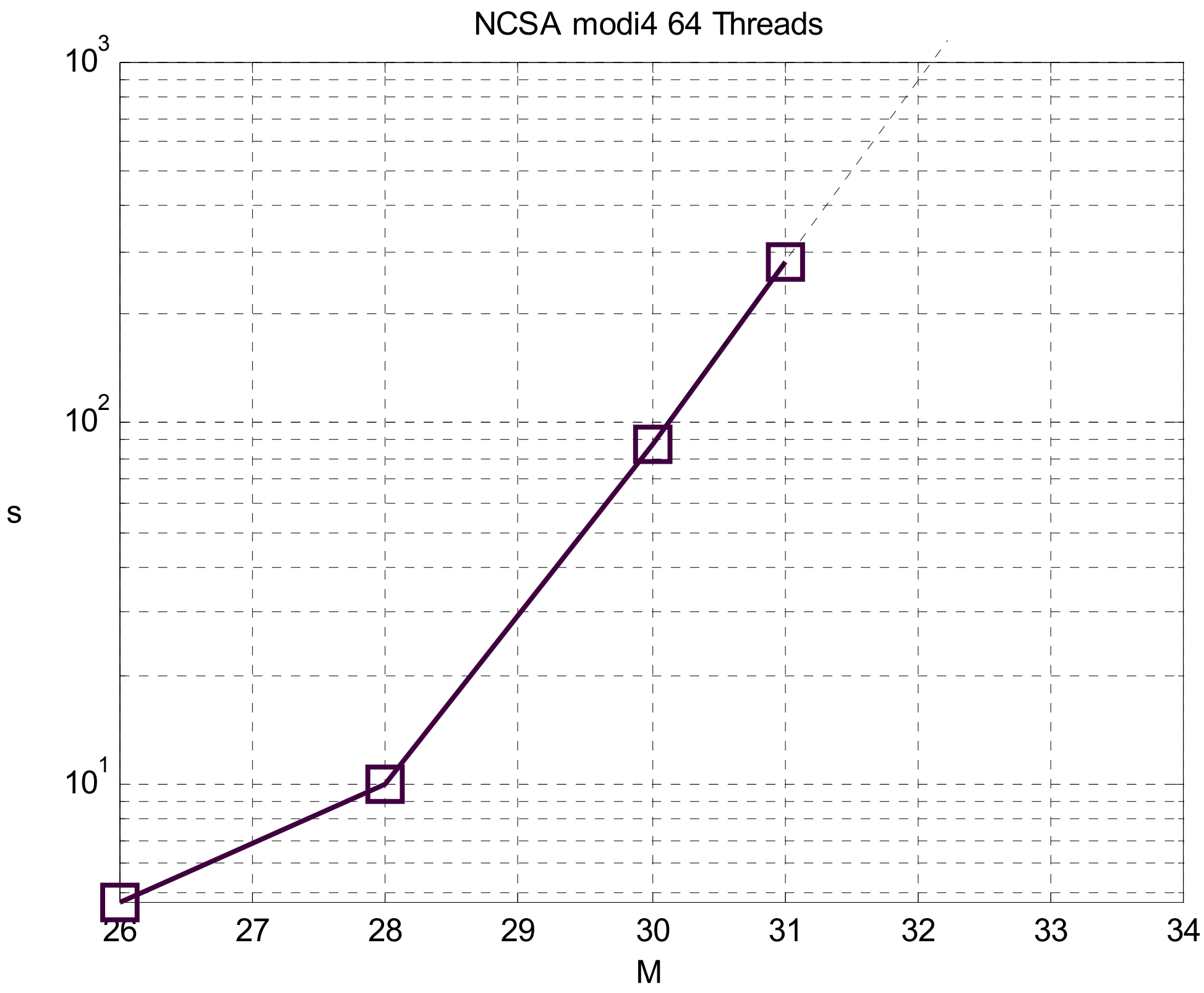

**Figure 4. Data Using up to 64 Threads**

## 5. CONCLUSIONS

Currently there is no other way to exercise non-trivial quantum algorithms except by simulation. Questions posed in the introduction have been answered at least in a special case. Without special programming, the modi4 classical parallel processor easily processed up to 32 qubits. This required less than 20 minutes. If parallelism can be exploited effectively, and data structures managed efficiently, it appears that many more qubits can be processed in the equivalent of a single modi4.

Compared to a basic accumulation method that avoids the unitary matrix, the sparse matrix approach has advantages, including the possibility of a user-controlled tradeoff between number of parallel processors and size of memory. The algorithm given can calculate a state vector from top to bottom, as seen, without special sorting. It has the advantage that it can save information about the unitary matrix, enabling new possibilities for design automation. Because a matrix approach is modifiable to accept quantum gates not envisioned today, this approach is well worth considering.

Future work involves finding ways to operate quantum computers themselves in parallel.

## Appendix 1
## CALCULATING THE INPUT STATE ON THE FLY

The input can be stored in the form of M qubits labeled $q_i$, $1 \le i \le M$, each represented as a pair of complex numbers $[q_i(0)\ q_i(1)]$. $\mathbf{k_1}$ is such that its elements are generated by a Kronecker (direct) matrix product:

$$\mathbf{k_1} = q_M \otimes q_{M-1} \otimes \ldots \otimes q_3 \otimes q_2 \otimes q_1$$

There is a way to obtain random $k_1(i)$. Imagine each qubit as being replace by |1>, that is, by the pair [0 1]. Then after a Kronecker product, $k_1(0)$ is addressed by M bits, all zero: 00…000, $k_1(1)$ is addressed by: 00…001, and so on up to $k_1(2^{M-1})$ which is addressed by M bits, all ones: 11…111.

If a particular element of $k_1(i)$ is wanted, simply use the binary form of i to identify the appropriate components of each qubit to be multiplied. For example, if $k_1(5)$ is wanted, use i = 101 to obtain $q_3(1)q_2(0)q_1(1)$.